\documentclass[11pt]{llncs}

\usepackage{amssymb}
\usepackage{amsmath}
\usepackage{amsthm}
\usepackage{textcomp}
\usepackage{array}
\usepackage[latin5]{inputenc}
\usepackage{lineno}
\usepackage{color}
\usepackage{comment}

\newcommand{\ket}[1]{|#1\rangle}
\newcommand{\braket}[2]{\langle #1|#2\rangle}
\newcommand{\cent}[0]{\mbox{\textcent}}
\newcommand{\dollar}[0]{\$}

\theoremstyle{definition}

\begin{document}
\title{NP has log-space verifiers with fixed-size public quantum registers}
\author{Abuzer Yakary{\i}lmaz$ ^{\mbox{\tiny 1,}} $\thanks{Yakary{\i}lmaz was
partially supported by T\"{U}B\.ITAK with grant 108E142 and FP7 FET-Open project QCS.} \and 
A. C. Cem Say$ ^{\mbox{\tiny 2,}} $\thanks{Say was partially supported by T\"{U}B\.ITAK with grant 108E142.}}

\institute{\scriptsize
$ ^{1} $University of Latvia, Faculty of Computing, Raina bulv. 19, R\={\i}ga, LV-1586, Latvia
\\
$ ^{2} $Bo\u{g}azi\c{c}i University, Department of Computer Engineering, Bebek 34342 \.{I}stanbul, Turkey
\\
\email{abuzer@lu.lv,say@boun.edu.tr}
}

\maketitle       
\begin{abstract}
	In classical Arthur-Merlin games, the class of languages whose membership proofs can be verified by Arthur using logarithmic space ($\mathsf{AM(log\mbox{-}space)}$)
	coincides with the class $\mathsf{P}$ \cite{Co89}. In this note, we show that if Arthur has a fixed-size quantum register 
	(the size of the register does not depend on the length of the input) 
	instead of another source of random bits, membership in any language in $ \mathsf{NP} $	
	can be verified with any desired error bound.
\end{abstract}

In a public-coin interactive proof system, a resource-bounded verifier (Arthur)  checks proofs presented to it in a sequence of messages by an all-powerful prover (Merlin), from which Arthur can hide no information. Arthur is supposed to accept all and only the correct proofs with high probability.  Quantum versions of these systems, where the verifier is a quantum computer, and  the messages consist of quantum bits, have also been examined, with the most general variant with polynomial bounds on the exchanged messages and the verifier runtime shown to be equivalent in computational power to the classical version \cite{MW05}. More restricted scenarios have also been examined; for instance, the class $\mathsf{QCMA}$ \cite{AN02,AK06}  corresponds to single-message quantum Arthur-Merlin games, where the proof is a classical string.\footnote{Watrous \cite{Wa09} proposes the usage of the name $\mathsf{MQA}$ for this class.}

In this paper, we restrict  this model further, by imposing a logarithmic space bound on the verifier, so that Arthur cannot even hold the entire proof string in memory, and take most of his ``quantumness" away, by limiting him to use  a  quantum register (with a state set of size just three) as his only source of randomness. Without that small register, Arthur is a deterministic logspace machine, and the class of languages that it can verify membership in is just $ \mathsf{NL} $. In the version where Arthur is allowed to use a classical random number generator, this class is known to equal $ \mathsf{P} $ \cite{Co89}. We will show that our version of Arthur with the small quantum register can verify membership in every language in $ \mathsf{NP} $.

We use the abbreviation qTM, with a lowercase q, for the quantum Turing machine model with which we represent the verifier, to stress that it uses only  a fixed number of quantum bits. The classical state set of Arthur is partitioned to three subsets, the reading states, communication states, and halting states. Every computational step of  the qTM consists of two stages: 
1) A selective quantum operation depending on the current classical state is performed on the quantum register, and
2) A deterministic transition updates the classical state, worktape content, and tape head positions depending on the current classical state $ s $, the currently scanned tape symbols, the outcome of the quantum operation in the first stage, and, 
if $ s $ is a communication state, the next proof symbol transmitted by Merlin.\footnote{The knowledgeable reader will have noticed that this model is an intermediate between 2-way finite automata with quantum and classical states \cite{AW02} and the Turing machines with both classical and quantum tapes of \cite{Wa03}.}

Merlin is assumed to learn the outcomes of the quantum operations performed by Arthur simultaneously with Arthur. Since these operations are the only source of randomness for Arthur, Merlin has complete information about Arthur's overall state at all times.

Let $ \mathsf{qAM(log\mbox{-}space)} $ be the class of languages that can be proven with bounded error to such verifiers that are restricted to use  $O(\log n)$ space on their classical work tape for an input string of length $n$. We prove that $ \mathsf{NP} \subseteq \mathsf{qAM(log\mbox{-}space)} $ as follows:

The well-known $ \mathsf{NP} $-complete language SUBSET-SUM is the collection of all strings of the form $ S \# a_{1} \#  \ldots \# a_{n} \# $
such that $ S $ and the $ a_{i} $'s are numbers in binary ($ 1 \le i \le n $),
and there exists a set $ I \subseteq \{1,\ldots,n\} $ satisfying $ \sum_{i \in I} a_{i} = S $, where $ n>0 $. For any language $L \in $ $ \mathsf{NP} $,  Arthur repeatedly computes the logspace reduction from $L$ to SUBSET-SUM in an infinite loop, and Merlin indicates which of the $ a_{i} $'s are the members of $I$.  During each iteration of this loop, Arthur restarts this process with a probability close to 1. With a great portion of the remaining (exponentially small) probability, Arthur subtracts the numbers indicated by Merlin from $S$, using his quantum register, as will be described below, and rejects if  $ \sum_{i \in I} a_{i} \neq S $. With the remaining (even smaller) probability, he accepts the proof.

In a recent paper \cite{YS10B}, Yakary{\i}lmaz and Say showed that real-time quantum finite automata (qfa's) can perform low-probability arithmetic calculations as the one described above. Arthur uses its quantum register to implement the corresponding qfa, which it repeatedly feeds with the corresponding substrings of the output of the reduction it is computing.

Membership in $L$ can be proven to the resulting verifier with one-sided error in exponential expected time. In our detailed report, we show how to tune the amplitudes of the quantum transitions to achieve the desired error bound.

\section*{TECHNICAL DETAILS}

\subsection*{Background} \label{app:background}

A random number generator can be viewed as a system that consists of an $n$-state \textit{probabilistic register} with state set 
$ Q = \{ q_{i} \mid 1 \leq i \leq n \} $, on which some stochastic operations (matrices) are applied, where $ n>0 $.
The distribution of the states of the register is represented by a stochastic column vector $ v $.
After applying a stochastic operator $ A $ on $ v $, observation of the register to learn the index of its current state yields the outcome ``$ i $" with probability $ p_i = v'[i] $, where $ v'=Av $.

A \textit{quantum register} (or any finite-dimensional quantum system) is an $ n $-dimensional 
Hilbert space, a complex vector space with inner product, denoted as $ \mathcal{H}_{n} $, where $ n>0 $.
The set $ \mathcal{B}_{n} = \{ \ket{q_{i}} \mid 1 \le i \le n \} $ is an orthonormal basis for $ \mathcal{H}_{n} $,
where the $ i^{th} $ entry of $ \ket{q_{i}} $ is 1 and the remaining entries are zeros.
Any \textit{quantum state} of the system is described by its \textit{state vector}, say $ \ket{\psi} $,
that is a linear combination of \textit{basis states}\footnote{We fixed it as $ \mathcal{B}_{n} $. 
However, note that, one can also select any other orthonormal basis.}
$ \ket{\psi} = \alpha_{1} \ket{q_{1}}+ \cdots + \alpha_{n} \ket{q_{n}} $,
where the number $ \alpha_{i} $ is the \textit{amplitude} of $ \ket{q_{i}} $, whose modulus squared ($ |\alpha_{i}|^{2} $)
gives the probability of being in state $ q_{i} $,
and $ \sum_{i} |\alpha_{i}|^{2}=1 $  ($ 1 \le i \le n $).
When $ \ket{\psi} $ contains more than one basis state with nonzero amplitude,
the system is said to be in a \textit{superposition} (of the basis states).

The most general operator applied to a quantum system is a superoperator,
which generalizes unitary and stochastic operators and also includes measurement.
Formally, a superoperator $ \mathcal{E} $ is composed by a finite number of operation elements,
$ \mathcal{E} = \{ E_{1}, \ldots, E_{k} \} $, satisfying that
\begin{equation}
	\label{eq:completeness}
	\sum_{i=1}^{k} E_{i}^{\dagger} E_{i} = I,
\end{equation}
where $ k \in \mathbb{Z}^{+} $ and the indices are the measurement outcomes.
When $\mathcal{E}$ is applied on a quantum system in state $\ket{\psi} $, i.e. $ \mathcal{E}(\ket{\psi}) $,
we obtain the measurement outcome $i$ with probability $ p_{i} = \braket{\tilde{\psi_{i}}}{\tilde{\psi_{i}}} $,
where $\ket{\tilde{\psi_{i}}}$ is calculated as $ \ket{\tilde{\psi}_{i}} = E_{i} \ket{\psi} $ and $1 \leq i \leq k$.
If the outcome $i$ is observed ($p_{i} > 0 $), the new state of the system 
is obtained by normalizing $ \ket{\tilde{\psi}_{i}} $, 
which is $ \ket{\psi_{i}} = \frac{\ket{\tilde{\psi_{i}}}}{\sqrt{p_{i}}} $.

Moreover, as a special operator, the quantum register can be initialized to a predefined quantum state.
We call this operator as \textit{the initialize operator}, denoted $ \acute{\mathcal{E}} $, which has only one outcome.

In this paper, we assume that the entries of stochastic and quantum operators are defined by rational numbers.

\subsection*{Definition of proof systems} \label{app:definition}

We give the definition of proof systems based on \cite{Co93A}. 
We call our new system \textit{qArthur-Merlin}, or shortly $ \mathsf{qAM} $.

An $ \mathsf{AM} $ (resp., a $ \mathsf{qAM} $)  system consists of a prover Merlin and a verifier Arthur (resp., qArthur).
Both Arthur and qArthur are resource-bounded Turing machines having 
a read-only input tape and a read/write work tape.
Each tape head has a two-way access.
As a source of branching,
the Arthur has a random number generator and
the qArthur has a (finite-size) quantum register instead.
The states of both are partitioned into reading, communication, and halting (accepting or rejecting) states
and both have a special communication cell that allows them to communicate with Merlin,
where the capacity of the cell is finite.

The one-step transitions of the verifiers can be described as follows.
\begin{itemize}
	\item When Arthur (resp., qArthur) is in a reading state:
		\begin{enumerate}
			\item Arthur applies a stochastic operator on its random number generator
				(resp., qArthur applies a quantum operator on its finite register) 
				based on the state and the symbols under the tape heads. 
				Then, the outcome is automatically written to communication cell.
			\item The verifier determines the next configuration of the verifier, 
				based on the symbol under the tape heads, the state and the outcome.
		\end{enumerate} 	
	\item When the verifier is in a communication state:
		\begin{enumerate}
			\item The verifiers writes a symbol on the communication cell with respect to the current state.
			\item Then, in response, Merlin writes a symbol in the cell.
			\item Based on the state and the symbol written by Merlin, the verifier
				defines the next state of the verifier.
		\end{enumerate}
\end{itemize}
Note that, the input is accepted or rejected when Arthur enters a accepting or rejecting states, respectively.

For a given input $ w $, the probability that $ (P,V) $ accepts (rejects) $ w $ is the 
cumulative accepting (rejecting) probabilities taken over all branches of the verifier.
The prover-verifier pair $ (P,V) $ is an $ \mathsf{AM} $ (or a $ \mathsf{qAM} $) proof system
with error probability $ \epsilon < \frac{1}{2} $ if
\begin{enumerate}
	\item for all $ w \in L $, the probability that $ (P,V) $ accepts $ w $ is greated than $ 1-\epsilon $,
	\item for all $ w \notin L $, and all provers $ P^* $, the probability that $ (P^*,V) $ rejects $ w $
		is greater than $ 1-\epsilon $.
\end{enumerate}

\subsection*{Notation} \label{app:notation}

$ \mathsf{AM(\mathfrak{restriction})} $ (or $ \mathsf{qAM(\mathfrak{restriction})} $)
is the class of languages which have a proof system such that Arthur (or qArthur) has the restrictions
denoted by $ \mathfrak{restriction} $.
We use restrictions on work space and specifically focus on constant space (denoted $ \mathsf{1} $)
or logarithmic space (denoted $\mathsf{log\mbox{-}space} $).
Note that, any TM with constant space can be converted to a finite state automaton.
So the verifier becomes a (two-way) finite state automaton. 
Note that, Arthur (resp., qArthur) becomes a two-way probabilistic finite automaton, 2pfa, \cite{Fr81} 
(resp., a two-way finite automaton with quantum and classical states, 2qcfa, \cite{AW02,YS11A}) 
if we remove the communication part.

\subsection*{SUBSET-SUM $\in \mathsf{qAM}(1) $} \label{app:subset-sum}

We present a $ \mathsf{qAM}(1) $ proof system for the $\mathsf{NP}$-complete language SUBSET-SUM, where qArthur is restricted with constant space.

SUBSET-SUM is the collection of all strings of the form $ S \# a_{1} \#  \ldots \# a_{n} \# $
such that $ S $ and the $ a_{i} $'s are numbers in binary ($ 1 \le i \le n $),
and there exists a set $ I \subseteq \{1,\ldots,n\} $ satisfying $ \sum_{i \in I} a_{i} = S $, where $ n>0 $.

At the beginning of the computation, the input is deterministically checked to see if it is of the form 
\begin{equation*}
	\left( \{0,1\}^{+} \# \right)\left( \{0,1\}^{+} \# \right)^{+}.
\end{equation*}
If not, it is rejected.
In the remaining part, we assume the input to be of the form  
\begin{equation*}
	S \# a_{1} \#  \ldots \# a_{n} \#,
\end{equation*}
where $ S $, the $ a_{i} $'s are numbers in binary ($ 1 \le i \le n $), and $ n>0 $.

The main idea is that qArthur scans the input from left to right in an infinite loop
and firstly encodes $ S $, and then subtracts the encoding of each of the  $ a_i $'s selected by Merlin, 
in some amplitudes of the classical states on the quantum register.
And at the end of the loop, qArthur tests whether the result is zero or not as described later.
Since our encoding procedure works by reducing the amplitude with a constant in each step,
the process can successfully be ended with a exponentially small probability depending on the length of the input.
Therefore, the loop is repeated with high probability.
The technical details are given below.

The register has 3 classical states, i.e. $ \{ q_{1}, q_2, q_{3} \} $.
We can divide the procedure into 5 parts.
The input head is moved right whenever the outcome ``$ \rightarrow $" is observed.
If the outcome ``A" or ``R" is observed, the input is accepted or rejected, respectively.
Otherwise, the procedure is restarted.
\begin{enumerate}
	\item The finite register is initialized on symbol $ \cent $:
		\begin{equation*}
			\ket{\psi_{0}} =
			\left( 
				\begin{array}{c}
					1 \\ 0 \\ 0 
				\end{array}
			\right).
		\end{equation*}
	\item The first binary integer ($ S $) is encoded into the amplitudes of $ \ket{q_{2}} $:
		$ \mathcal{E}_{\sigma} $ is applied on the quantum register when reading $ \sigma \in \{0,1\} $,
		i.e. 
		\begin{equation*}			
			\mathcal{E}_{0} = 
			\left\lbrace
				\underbrace{
				\frac{1}{3}\left( 
				\begin{array}{rrr}
					~~1 & ~~0 & ~~0 \\
					0 & 2 & 0 \\
					0 & 0 & 1  \\
				\end{array}
				\right)
				}_{\rightarrow};
				~
				\underbrace{
				\frac{1}{3}\left( 
				\begin{array}{rrr}				
					~~2 & ~~0 & -2  \\
					2 & 0 & 2 \\
					0 & 2 & 0 \\
				\end{array}
				\right)
				}_{\circlearrowright};
				~
				\underbrace{
				\frac{1}{3}\left( 
				\begin{array}{rrr}	
					~~0 & ~~1 & ~~0 \\
					0 & 0 & 0 \\
					0 & 0 & 0 \\
				\end{array}
			\right)
			}_{\circlearrowright}
			\right\rbrace
		\end{equation*}
		and
		\begin{equation*}			
			\mathcal{E}_{1} = 
			\left\lbrace
				\underbrace{
				\frac{1}{3}\left( 
				\begin{array}{rrr}
					~~1 & ~~0 & ~~0 \\
					1 & 2 & 0  \\
					0 & 0 & 1  \\
				\end{array}
				\right)
				}_{\rightarrow};
				~
				\underbrace{
				\frac{1}{3}\left( 
				\begin{array}{rrr}				
					~~2 & -1 & ~~0 \\
					1 & 0 & 2 \\
					1 & 0 & -2 \\
				\end{array}
				\right)
				}_{\circlearrowright};
				~
				\underbrace{
				\frac{1}{3}\left( 
				\begin{array}{rrr}	
					~~1 & ~~0 & ~~0 \\
					0 & 2 & 0 \\
					0 & 0 & 0 \\
				\end{array}
			\right)
			}_{\circlearrowright}
			\right\rbrace.
		\end{equation*}
		$ \mathcal{E}_{\#} $ is applied on the quantum register when reading $ \# $, i.e.
		\begin{equation*}
			\mathcal{E}_{\#} = \left\lbrace
				\underbrace{\frac{1}{3} \mathcal{I}}_{\rightarrow}; ~ 
				\underbrace{\frac{2}{3} \mathcal{I}}_{\circlearrowright}; ~ 
				\underbrace{\frac{2}{3} \mathcal{I}}_{\circlearrowright}
			\right\rbrace.
		\end{equation*}
	\item Each $ a_{i} $  ($ 1 \leq i \leq n $) is encoded into the amplitude of $ \ket{q_{3}} $:
		$ \mathcal{E}_{\sigma}' $ is applied on the quantum register when reading $ \sigma \in \{0,1\} $,
		i.e. 
		\begin{equation*}			
			\mathcal{E}_{0}' = 
			\left\lbrace
				\underbrace{
				\frac{1}{3}\left( 
				\begin{array}{rrr}
					~~1 & ~~0 & ~~0 \\
					0 & 1 & 0 \\
					0 & 0 & 2 \\
				\end{array}
				\right)
				}_{\rightarrow};
				~
				\underbrace{
				\frac{1}{3}\left( 
				\begin{array}{rrr}				
					~~2 & ~~2 & ~~0 \\
					2 & -2 & 0 \\
					0 & 0 & 2 \\
				\end{array}
				\right)
				}_{\circlearrowright};
				~
				\underbrace{
				\frac{1}{3}\left( 
				\begin{array}{rrr}	
					~~0 & ~~0 & ~~1 \\
					0 & 0 & 0 \\
					0 & 0 & 0 \\
				\end{array}
			\right)
			}_{\circlearrowright}
			\right\rbrace
		\end{equation*}
		and
		\begin{equation*}			
			\mathcal{E}_{1}' = 
			\left\lbrace
				\underbrace{
				\frac{1}{3}\left( 
				\begin{array}{rrr}
					~~1 & ~~0 & ~~0 \\
					0 & 1 & 0 \\
					1 & 0 & 2 \\
				\end{array}
				\right)
				}_{\rightarrow};
				~
				\underbrace{
				\frac{1}{3}\left( 
				\begin{array}{rrr}				
					~~2 & 0 & -1 \\
					1 & 2 & ~~0 \\
					1 & -2 & 0 \\
				\end{array}
				\right)
				}_{\circlearrowright};
				~
				\underbrace{
				\frac{1}{3}\left( 
				\begin{array}{rrr}	
					~~1 & ~~0 & ~~0 \\
					0 & 0 & 2 \\
					0 & 0 & 0 \\
				\end{array}
			\right)
			}_{\circlearrowright}
			\right\rbrace.
		\end{equation*}
	\item If an $ a_{i} $ ($ 1 \leq i \leq n $) is \textit{selected} on Merlin's advice received on symbol $ \# $, 
		it is subtracted from 	the number represented by the amplitude of $ q_{2} $:
		$ \mathcal{E'}_{\#} $ is applied on the quantum register, i.e.
		\begin{equation*}	\small \mspace{-20mu}		
			\mathcal{E}_{\#}' = 
			\left\lbrace
				\underbrace{
				\frac{1}{3}\left( 
				\begin{array}{rrr}
					~~1 & ~~0 & ~~0 \\
					0 & 1 & -1 \\
					0 & 0 & 0 \\
				\end{array}
				\right)
				}_{\rightarrow};
				~
				\underbrace{
				\frac{1}{3}\left( 
				\begin{array}{rrr}				
					~~0 & -1 & ~~1 \\
					2 & 1 & -1 \\
					2 & -1 & 1 \\
				\end{array}
				\right)
				}_{\circlearrowright};
				~
				\underbrace{
				\frac{1}{3}\left( 
				\begin{array}{rrr}	
					~~0 & ~~2 & ~~2 \\
					0 & 0 & 0 \\
					0 & 0 & 0 \\
				\end{array}
				\right)
				}_{\circlearrowright};
				~
				\underbrace{
				\frac{1}{3}\left( 
				\begin{array}{rrr}	
					~~0 & ~~1 & ~~0 \\
					0 & 0 & 1 \\
					0 & 0 & 0 \\
				\end{array}
				\right)
				}_{\circlearrowright}			
			\right\rbrace.
		\end{equation*}
		If it is not selected, $ \mathcal{E''}_{\#} $ is applied on the quantum register, i.e.
		\begin{equation*}			
			\mathcal{E}_{\#}' = 
			\left\lbrace
				\underbrace{
				\frac{1}{3}\left( 
				\begin{array}{rrr}
					~~1 & ~~0 & ~~0 \\
					0 & 1 & 0 \\
					0 & 0 & 0 \\
				\end{array}
				\right)
				}_{\rightarrow};
				~
				\underbrace{
				\frac{1}{3}\left( 
				\begin{array}{rrr}				
					~~2 & -2 & ~~0 \\
					2 & 2 & 0 \\
					0 & 0 & 3 \\
				\end{array}
				\right)
				}_{\circlearrowright}			
			\right\rbrace.
		\end{equation*}
		Note that, the amplitude of $ \ket{q_{3}} $ is set to 0 after each of these transformations.
	\item The decision is given on $ \dollar $:
	$ \mathcal{E}_{\dollar} $ is applied on the quantum register when reading $ \dollar $,
		i.e.
		\begin{equation*}			
			\mathcal{E}_{\dollar} = 
			\left\lbrace
				\underbrace{
				\frac{1}{3}\left( 
				\begin{array}{rrr}
					~~1 & ~~0 & ~~0 \\
					0 & 0 & 0 \\
					0 & 0 & 0 \\
				\end{array}
				\right)
				}_{A};
				~
				\underbrace{
				\frac{1}{3}\left( 
				\begin{array}{rrr}				
					~~0 & ~~0 & ~~0 \\
					0 & 3 & 0 \\
					0 & 0 & 0 \\
				\end{array}
				\right)
				}_{R};
				~
				\underbrace{
				\frac{1}{3}\left( 
				\begin{array}{rrr}	
					~~2 & ~~0 & ~~0 \\
					2 & 0 & 0 \\
					0 & 0 & 3 \\
				\end{array}
			\right)
			}_{\circlearrowright}
			\right\rbrace.
		\end{equation*}
\end{enumerate}
Let $ w $ be the input. Then, the state of quantum register before reading $ \dollar $ is
\begin{equation*}
	\ket{\psi_{|w|}} = \left( \frac{1}{3} \right)^{|w|} 
	\left( \begin{array}{c}
					1 \\ S-T \\ 0
				\end{array} \right),
\end{equation*}
where $ T $ is the cumulative sum of selected $ a_{i} $'s  by Merlin.
After applying $ \mathcal{E}_{\dollar} $, the input is rejected with probability
\begin{equation*}
	\left( \frac{1}{3} \right)^{2|w|+2} 
	\left( 3 S -3 T \right)^{2},
\end{equation*}
which is at least $ 9 \left( \frac{1}{3} \right)^{2|w|+2} $ if $ S \neq T $
and is exactly equal to 0 if $ S=T $.
On the other hand, it is always accepted with probability $ \left( \frac{1}{3} \right)^{2|w|+2} $.
Therefore, if the input is a member of SUBSET-SUM, there exists a Merlin such that it is accepted exactly.
On the other hand, if the input is not a member of SUBSET-SUM,
whatevet Merlin says, it is rejected with a probability at least $ \frac{9}{10} $.
The error bound can be  reduced easily by using conventional probability amplification techniques.

\subsection*{$\mathsf{NP}\subseteq \mathsf{qAM}( \mathsf{log\mbox{-}space})$} \label{app:concl}

Since any language $ L \in \mathsf{NP} $ is log-space reducible \cite{Pa94} 
to SUBSET-SUM, we conclude that
Arthur can verify membership in any such $ L $ with high probability of correctness using only  logarithmic (classical) space, and a fixed-size quantum register to implement the qfa described above.

\bibliographystyle{alpha}
\bibliography{YakaryilmazSay}

\end{document}